\documentclass[a4paper,11pt]{article}
\pdfoutput=1 

\usepackage{jinstpub} 
\usepackage{graphicx} 
\usepackage{url}

\title{CYGNO: a gaseous TPC with optical readout for dark matter directional search}

\author[a,b]{E. Baracchini,}
\author[c]{L. Benussi,}
\author[c]{S. Bianco,}
\author[c]{C. Capoccia,}
\author[c,d]{M. Caponero,}
\author[e,f]{G. Cavoto,}
\author[a,b]{A. Cortez,}
\author[e]{I. A. Costa,}
\author[e]{E. Di Marco,}
\author[e,1]{G. D'Imperio,\note{Corresponding author.}}
\author[a,b]{G. Dho,}
\author[e]{F. Iacoangeli,}
\author[c]{G. Maccarrone,}
\author[e,h]{M. Marafini,}
\author[c]{G. Mazzitelli,}
\author[e,f]{A. Messina,}
\author[g]{R. A. Nobrega,}
\author[c]{A. Orlandi,}
\author[c]{E. Paoletti,}
\author[c]{L. Passamonti,}
\author[i,j]{F. Petrucci,}
\author[c]{D. Piccolo,}
\author[c]{D. Pierluigi,}
\author[e]{D. Pinci,}
\author[e]{F. Renga,}
\author[c]{F. Rosatelli,}
\author[c]{A. Russo,}
\author[c,k]{G. Saviano,}
\author[c]{and S. Tomassini}
\affiliation[a]{Gran~Sasso~Science~Institute~L'Aquila, I-67100, Italy}
\affiliation[b]{Istituto Nazionale di Fisica Nucleare, Laboratori Nazionali del Gran Sasso, Assergi L'Aquila, I-67100 Italy}
\affiliation[c]{Istituto Nazionale di Fisica Nucleare, Laboratori Nazionali di Frascati, Frascati, I-00040, Italy}
\affiliation[d]{ENEA Centro Ricerche Frascati, Frascati, I-00040, Italy}
\affiliation[e]{Istituto~Nazionale~di~Fisica~Nucleare Sezione di Roma, Roma, I-00185, Italy}
\affiliation[f]{Dipartimento di Fisica Sapienza Universit\`a di Roma, Roma, I-00185, Italy}
\affiliation[g]{Universidade Federal de Juiz de Fora, Juiz de Fora, Brazil}
\affiliation[h]{Museo Storico della Fisica e Centro Studi e Ricerche "Enrico Fermi", Roma, I-00184, Italy}
\affiliation[i]{Dipartimento di Matematica e Fisica, Universit\`a Roma TRE, Roma, I-00146, Italy}
\affiliation[j]{Istituto Nazionale di Fisica Nucleare, Sezione di Roma TRE, Roma, I-00146, Italy}
\affiliation[k]{Dipartimento di Ingegneria Chimica, Materiali e Ambiente, Sapienza Universit\`a di Roma, Roma, I-00185, Italy}

\emailAdd{giulia.dimperio@roma1.infn.it}

\abstract{
  The CYGNO project has the goal to use a gaseous TPC with optical readout to detect dark matter and solar neutrinos with low energy threshold and directionality.
The CYGNO demonstrator will consist of 1~m$^3$ volume filled with He:CF$_4$ gas mixture at atmospheric pressure.
Optical readout with high granularity CMOS sensors, combined with fast light detectors, will provide a detailed reconstruction of the event topology.
This will allow to discriminate the nuclear recoil signal from the background, mainly represented by low energy electron recoils induced by radioactivity. 
Thanks to the high reconstruction efficiency, 
CYGNO will be sensitive to low mass dark matter, and will have the potential to overcome the neutrino floor, that ultimately limits non-directional dark matter searches.
}

\keywords{Dark Matter detectors, Particle tracking detectors, Time projection chambers, Micropattern gaseous detectors, CMOS readout of gaseous detectors}

\proceeding{International Conference "Instrumentation for Colliding Beam Physics"\\
  24-28 February 2020\\
  Budker Institute of Nuclear Physics,Novosibirsk, Russia}

\begin{document}
\maketitle
\flushbottom

\section{Introduction}
\label{sec:intro}

Gas detectors are interesting candidates for dark matter search for several reasons.
\begin{itemize}
  
  \item Nuclear recoils, induced by dark matter (or neutrino) scattering, might travel in the gas enough to reconstruct the direction of the recoil, and therefore of the incoming particle. Directionality is a powerful handle to distinguish dark matter from neutrinos and to overcome the neutrino floor. In fact neutrinos comes mostly from the Sun, while dark matter, in Earth's reference system, is expected to come from the Cygnus constellation, that never matches the position of the Sun during the year.
 
  \item Track topology of the nuclear recoil signal is very different from the background, represented mostly by gamma-induced electron recoils due to environmental radioactivity. Therefore a good background rejection for energy greater of few~keV is expected.

  \item Gas mixtures containing large quantity of helium allow to reach low energy detection threshold, close to $\sim$1~keV, improving the sensitivity to low dark matter masses of the order of $\sim$1 GeV.

\end{itemize}
On the other hand, a gaseous target has the disadvantage of low density, therefore large volumes are needed to increase the target mass and lower the sensitivity. 

The most sensitive dark matter detectors today are based on ton-scale TPCs of noble liquids~\cite{Xenon,Darkside}. 
The possibility to scale up to very large masses pushes the sensitivity to low cross sections, but these techniques will be ultimately limited by the so called "neutrino floor".
Neutrino's scattering events will be in fact indistinguishable from dark matter. 
Low dark matter masses of O(1 GeV) are still largely unexplored and the best performance is now reached by cryogenic detectors~\cite{CRESST}, that are much more difficult to scale up to large dimensions. 

A gas TPC like CYGNO can give an important contribution to the dark matter quest. 
The introduction of directionality is in fact a fundamental tool to get over the neutrino background problem.
Moreover, the low detection threshold allows to compete with the most sensitive  detectors to low mass dark matter, with relatively small sensitive mass (few kg).  
For the same reasons, CYGNO is also a promising detector for solar neutrinos.

The choice for CYGNO detector is to use a He:CF$_4$ mixture with approximately 60:40 proportion at atmospheric pressure, a triple-GEM electron multiplier and a optical readout. 
The first phase of CYGNO, will consists of 1~m$^3$ of sensitive gas operated at Gran Sasso National Laboratories underground (LNGS), while in a second phase a larger volume of about $30-100$~m$^3$ is foreseen.
A parallel R\&D is also carried out to study a similar gas mixture with the add of a small quantity of SF$_6$ to obtain negative ion drift.
A recent study~\cite{nitec} demonstrated the possibility to operate a chamber with negative ion drift using He:CF$_4$:SF$_6$ gas mixture at nearly atmospheric pressure. 
The INITIUM project, recently funded with a ERC consolidator grant, is being developed in collaboration and synergy with CYGNO, in order to operate the detector also with negative ions as charge carriers.

\section{R\&D and results with LEMOn prototype}
\label{sec:optical_readout}

The CYGNO detector will use optical sensors to read the light produced at the electron multiplication stage. 
This idea is not new~\cite{opt-readout1,opt-readout2}, but the recent development of CMOS sensors makes it an interesting approach for a gaseous TPC with GEM electron multipliers.

There are a number of advantages using the optical readout.
\begin{itemize}
  
  \item Modern optical sensors provide high granularity with very low noise.

  \item They can be placed outside the detector (no interference with HV and reduced radioactive contamination close to the sensitive volume).
  
  \item Using suitable lens, large areas can be monitored with small sensors.
  
  \item Combining the CMOS 2D readout with a time measurement, using a fast light detector like a PMT or a SiPM, 3D reconstruction of the tracks can be achieved.

\end{itemize}

%

The feasibility of a TPC with He:CF$_4$ and optical readout has been already demonstrated with a small chamber and drift length of about 1~cm~\cite{orange}. 
The effect of electron diffusion in a larger detector (LEMOn, Large Elliptical Module with Optical readout) has also been extensively studied up to about 20~cm drift length~\cite{lemon}.

The LEMOn prototype has a 7~liter active drift volume, surrounded by an elliptical field cage ($200\times200\times240$ mm$^3$)
and a rectangular triple GEM structure of $200\times240$ mm$^2$, with transfer gaps among the GEMs of 2~mm.
The light produced by the GEMs is read by an ORCA-Flash~4.0 camera, placed 52~cm away from the last GEM and by a PMT on the cathode side.
The CMOS sensor of the camera has high granularity ($2048\times2048$ pixels), very low noise (around two photons per pixel), high sensitivity (70\% of QE @ 600~nm) and linear response.
The camera is instrumented with a Schneider lens (f$/0.95-25$~mm) and each pixel covers an effective area of $125 \times 125~\mu$m$^2$.
Some recent results with LEMOn detector are reported in this paper.

Typical operating conditions of the detector are: a drift electric field E$_d= 600$~V/cm, an electric field of V$_{GEM} = 460$~V for each GEM plane, and a transfer field between the GEMs of E$_t = 2$~kV/cm.

\subsection{Measurements with $^{55}$Fe source}
\begin{figure}[htbp]
\centering 
\includegraphics[width=.55\textwidth]{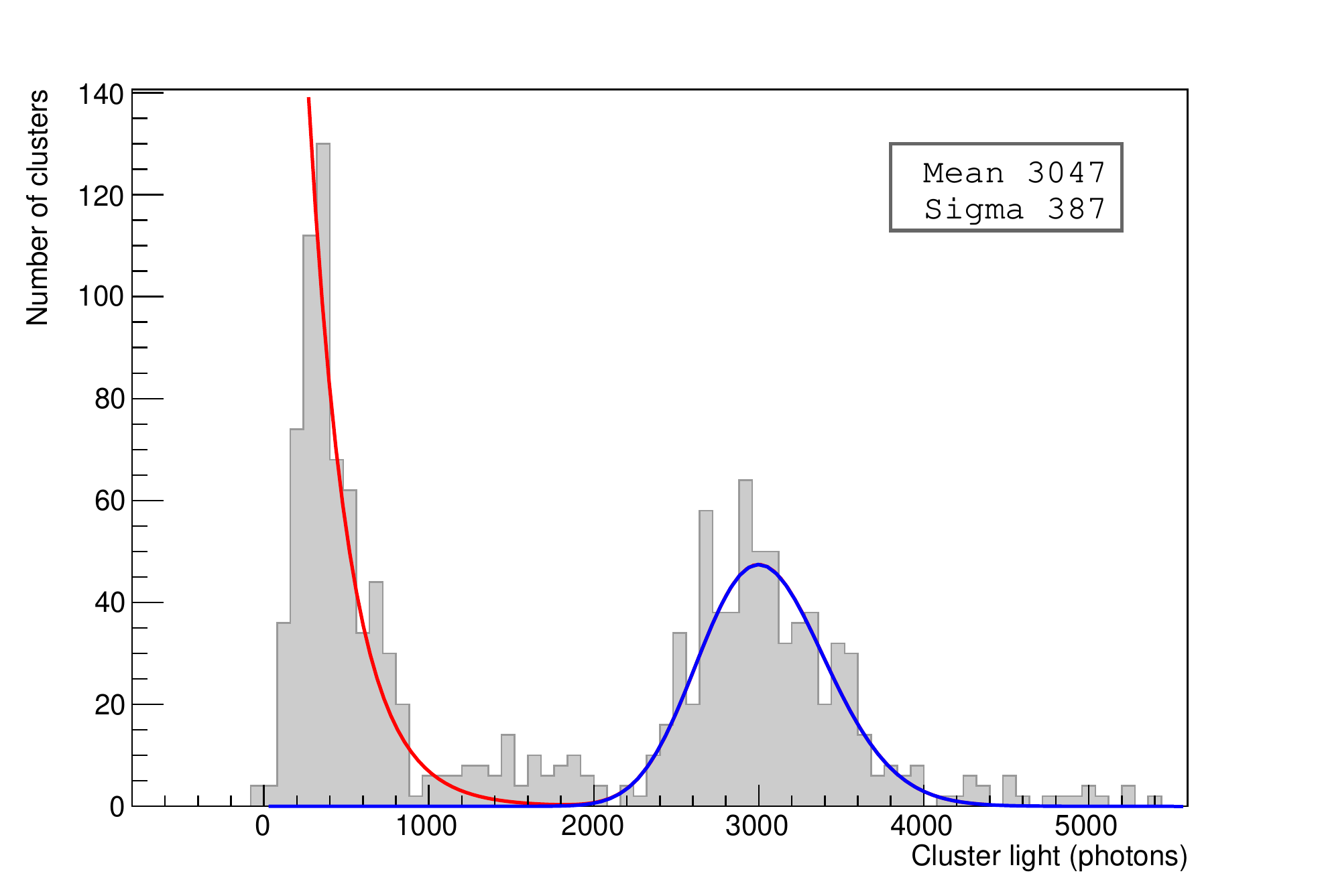}
  \caption{\label{fig:light_distr} Light distribution of clusters.} 
\end{figure}

The energy threshold has been studied exposing the detector to a $^{55}$Fe source that emits x-rays around 5.9~keV.
The $^{55}$Fe signal in LEMOn appears like light spots of about 2~mm diameter, due to the electron diffusion effect. 
Close pixels that are above a certain threshold, to avoid the electronic noise, are reconstructed as a cluster.
Fig.~\ref{fig:light_distr} shows the light distribution in the clusters: the gaussian shape in blue corresponds to the $^{55}$Fe signal. 
Looking at the background distribution in red it is possible to set an energy threshold around 2~keV. 
From the width of the gaussian it is possible to obtain the energy resolution, that is about 30\% FWHM at 5.9~keV. 

\subsection{Measurements with high energy electrons}
\begin{figure}[htbp]
\centering 
\includegraphics[width=.60\textwidth]{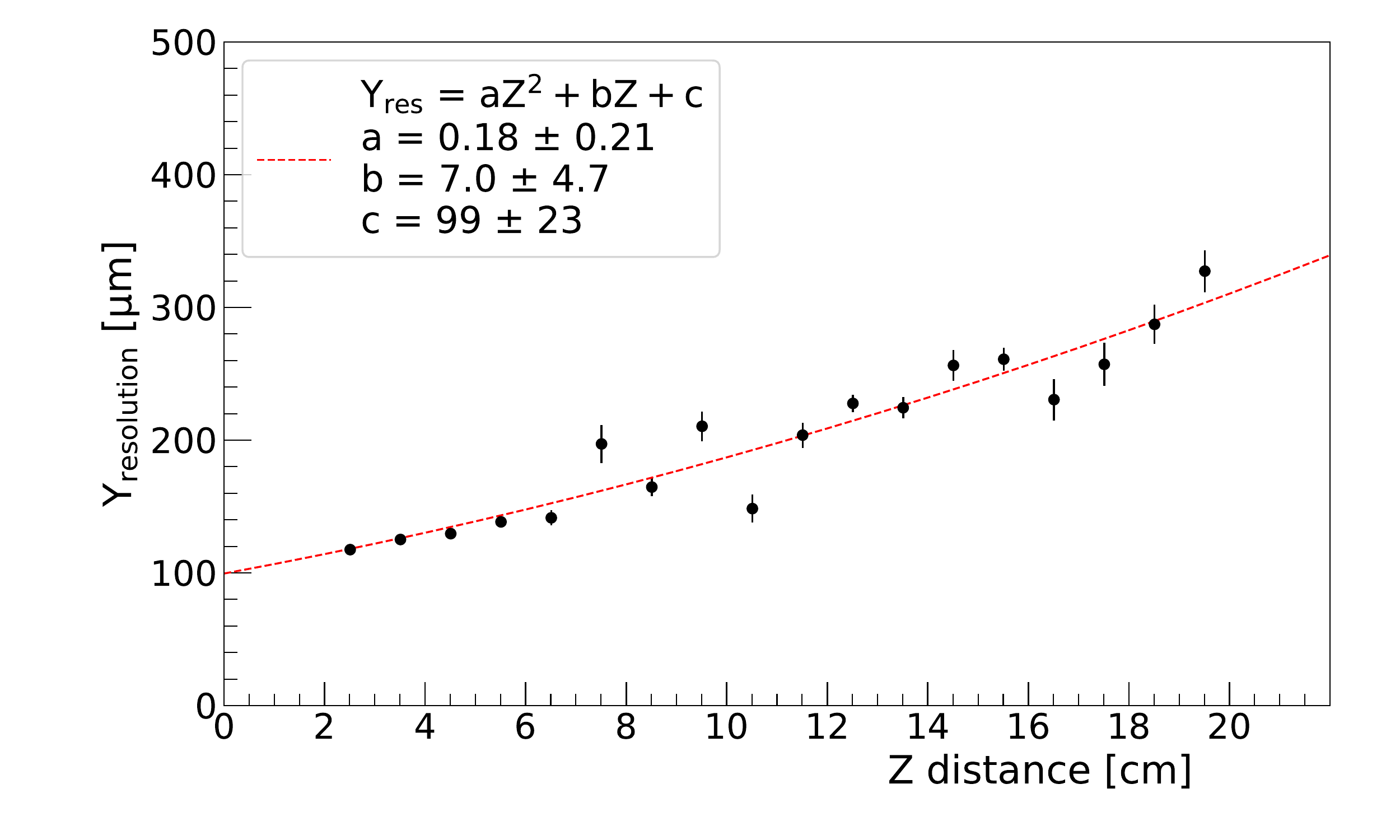}
  \caption{\label{fig:yres} Resolution on the transverse coordinate (Y) at different drift paths (Z), obtained with a drift field to 0.6kV/cm.}
\end{figure}

Tracking performances of LEMOn have been tested also with 450~MeV electrons at the Beam Test Facility in Frascati National Laboratories.
Studying tracks at different distance from the GEM plane, the effect of diffusion on the resolution was measured cutting the track in slices and measuring the RMS of the Y position (see Fig.~\ref{fig:yres}).
A resolution between 100 and 300 $\mu$m on the transverse coordinate is achieved for drift path ranging from 2 to 20~cm.

\begin{figure}[htbp]
\centering 
\includegraphics[width=.55\textwidth]{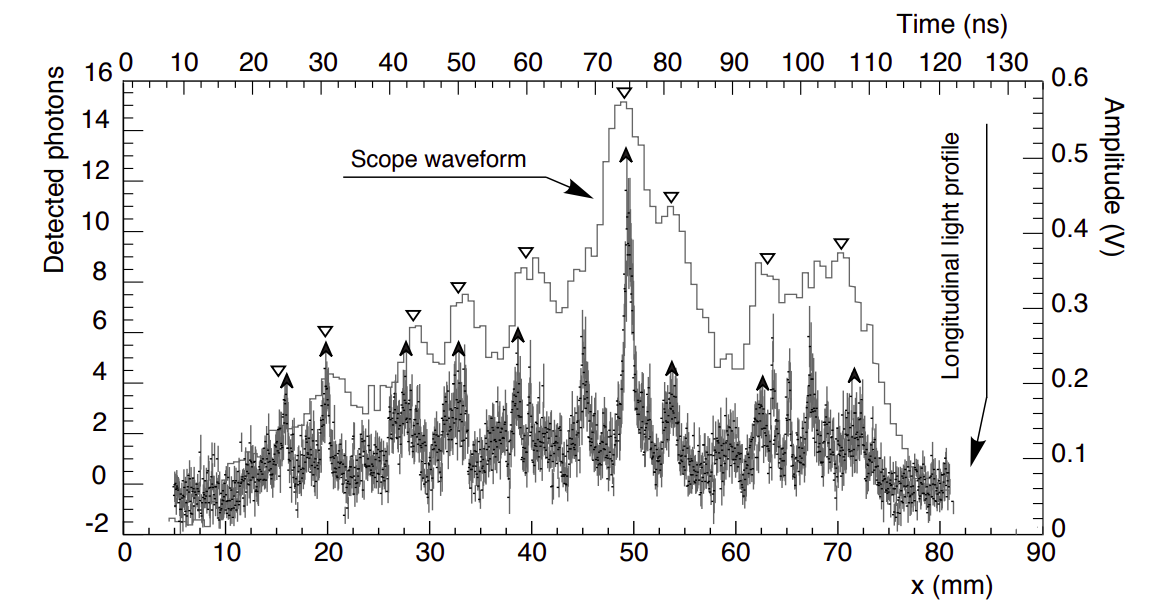}
\includegraphics[width=.40\textwidth]{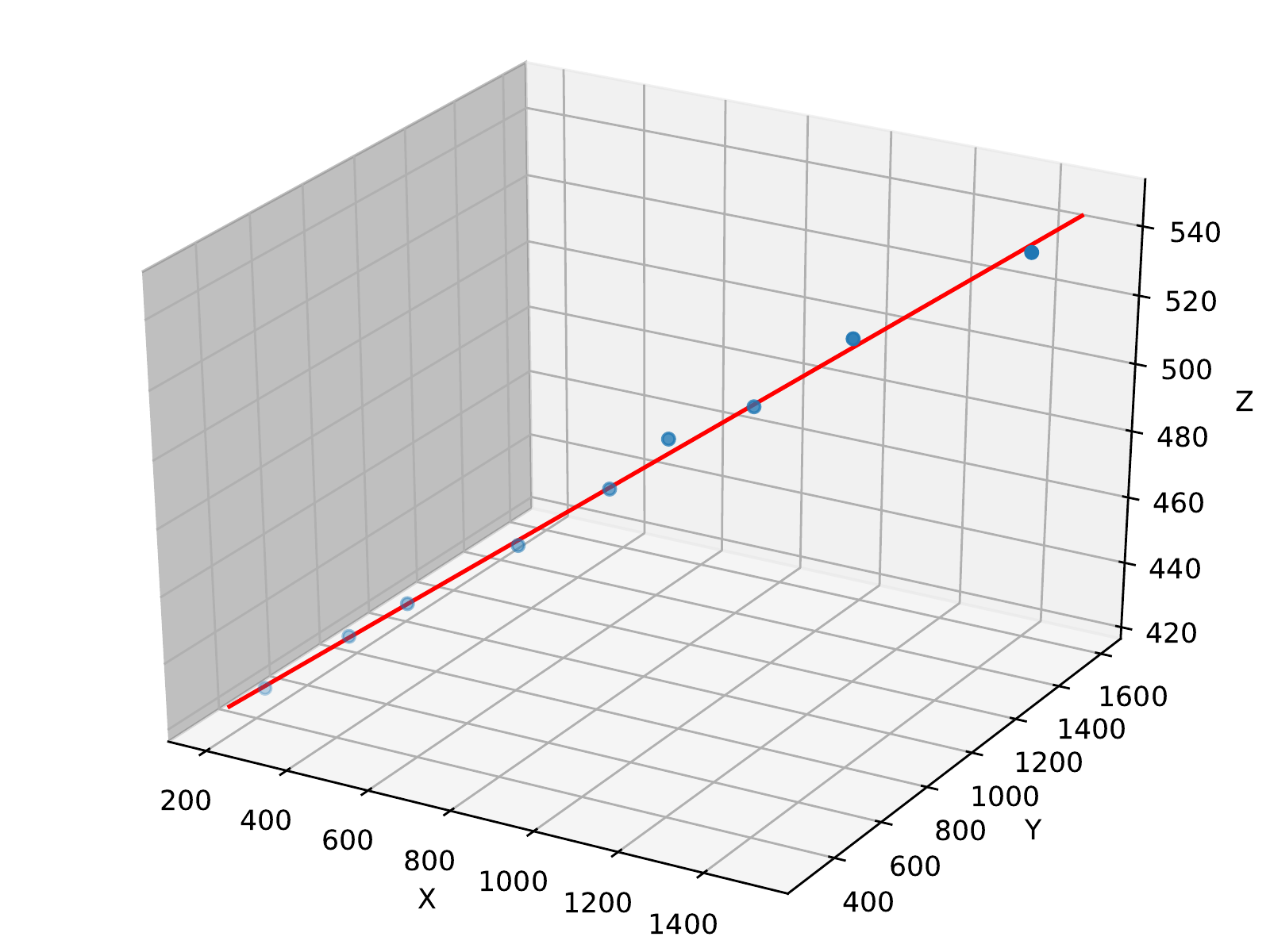}
  \caption{\label{fig:combined_readout} Lateral profile of the light detected by the CMOS sensor along with the waveform of the PMT signal for the same event (left) and example of a reconstructed 3D track (right).}
\end{figure}

When a track is tilted with respect to the detector readout plane, ionization clusters arrive to the GEMs at different times, with an average drift velocity of 7.2~cm/$\mu$s.
Typical CMOS exposure times are long (of the order of 10-100 ms), but using a fast light detector like a PMT, it is possible to measure the time of arrival of the clusters, thus providing a measurement of the third coordinate. Combination of the CMOS sensor and PMT allows a 3D reconstruction of the track with a resolution of 100~$\mu$m (see Fig.~\ref{fig:combined_readout}).

\subsection{Measurements with AmBe neutron source}
\begin{figure}[htbp]
\centering 
  \includegraphics[width=.30\textwidth]{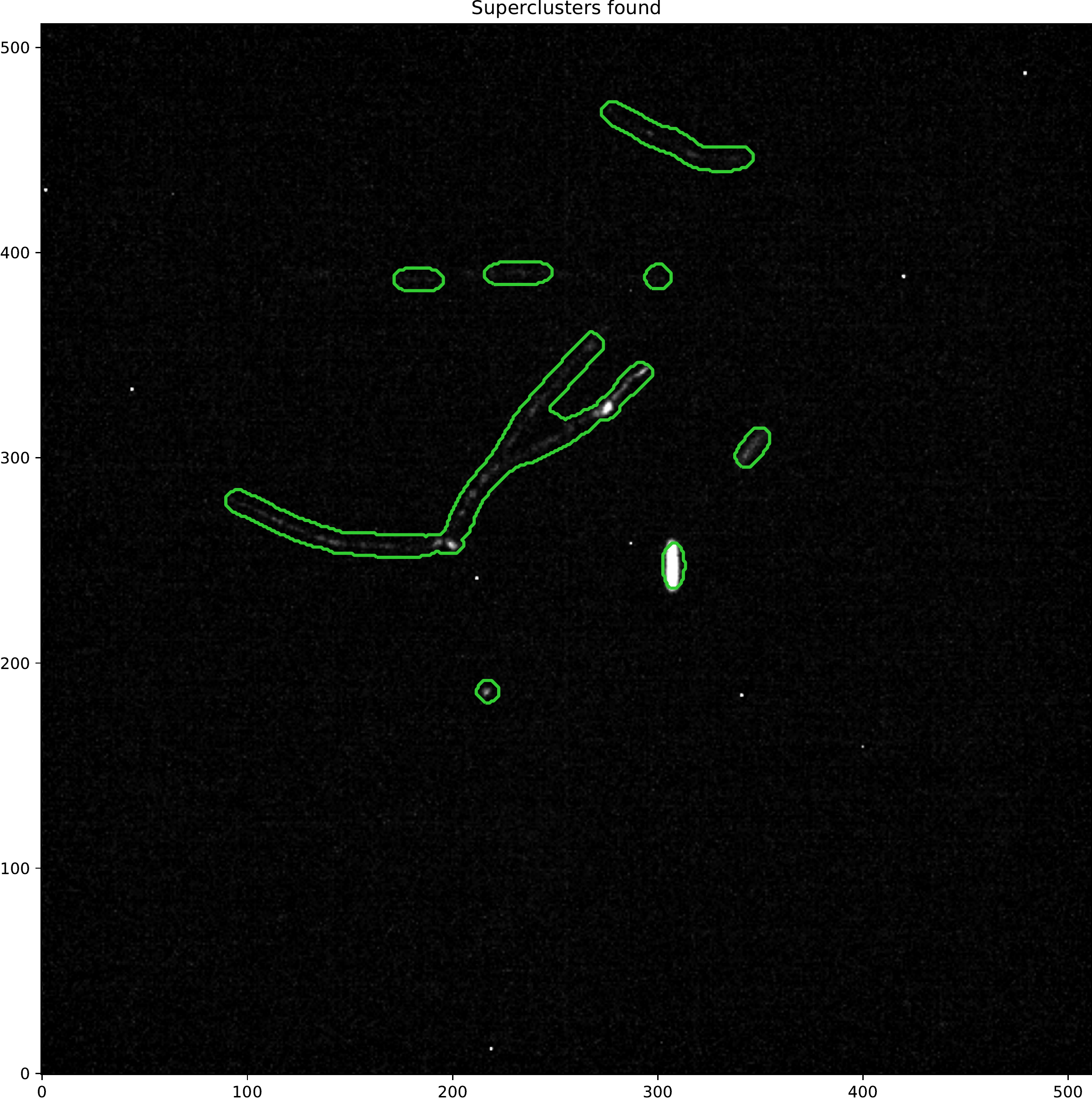}
\quad
  \includegraphics[width=.30\textwidth]{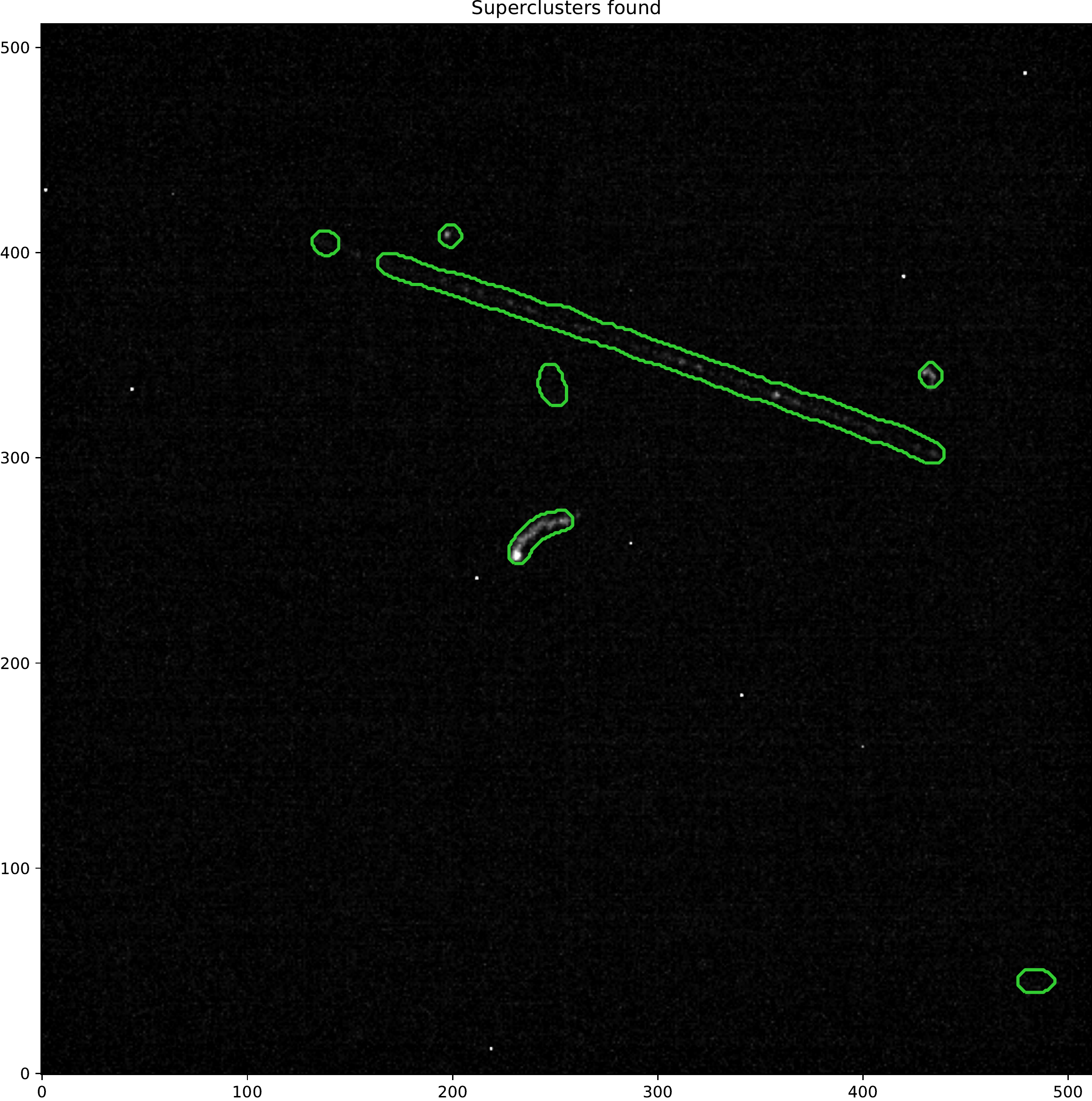}
\quad
  \includegraphics[width=.30\textwidth]{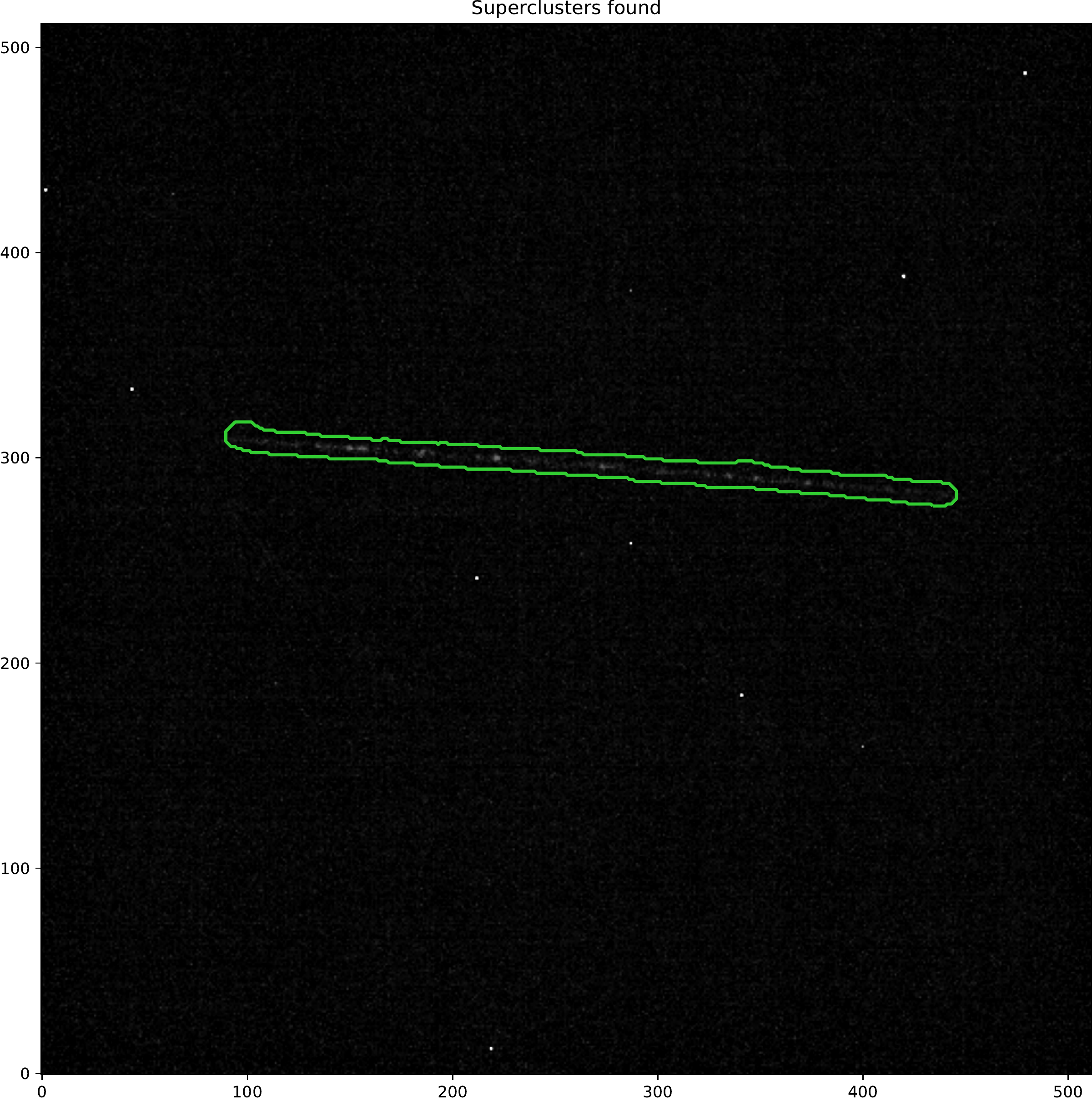}
  \caption{\label{fig:recoplot} Example of reconstruction of the tracks in LEMOn with AmBe source (left and middle picture), and without source (right picture).}
\end{figure}

In order to study the particle identification capability and the possibility to discriminate between signal from nuclear recoils and background from electron recoils, LEMOn was also tested with a neutron source of AmBe.
AmBe source provides neutrons with energy spectrum in the range 1-11~MeV; though it is not a pure source of neutrons, and also photons with an energy of 59.5~keV and 4.4~MeV are produced. 
Given the difference of the energy loss function for different particles and energies, a different topology of the tracks is expected.
For the identification of different particles, a reconstruction algorithm based on clustering of tracks with different light density has been developed. 
In Fig.~\ref{fig:recoplot} some examples of clustering on data with and without AmBe source are shown. 
The intense short track in the left picture is a nuclear recoil candidate of density 18.2 photons/pixel. 
The curve track in the middle picture is an example of another topology of tracks present in the AmBe data, with lower density (around 10-15 photons/pixel) and longer tracks, possibly the 59.5~keV photons. 
The long straight tracks present both in the middle and in the right picture (the latter from a data sample without source) represent the typical low density tracks (around 4-5 photons/pixel) due to cosmic rays.

%

\section{CYGNO detector design}
\label{sec:cygno}

In the phase-1, CYGNO detector  will have a cubic sensitive volume of 1~m$^3$, divided in two drift regions of 50~cm, separated by a central cathode.
The two readout planes will be equipped with triple-GEM stacks of 1 m$^2$ total surface, one on each side. 
A set of 18 cameras (9 per side) will look at a $33\times33$ cm$^2$ surface each. 
Fast light detectors (PMTs or SiPMs) will be also installed in between the cameras, looking at the GEMs, to allow the reconstruction of the third coordinate.
The technical design of the detector will be completed in 2020 and followed  by  the  construction,  with  the  goal  of  running  the experiment at LNGS in 2021.
The CYGNO phase-2 foresees a scaling up to higher volume, around $30-100$ m$^3$.


\section{Background simulations}
\label{sec:simulations}

\begin{figure}[htbp]
\centering 
\includegraphics[width=.60\textwidth]{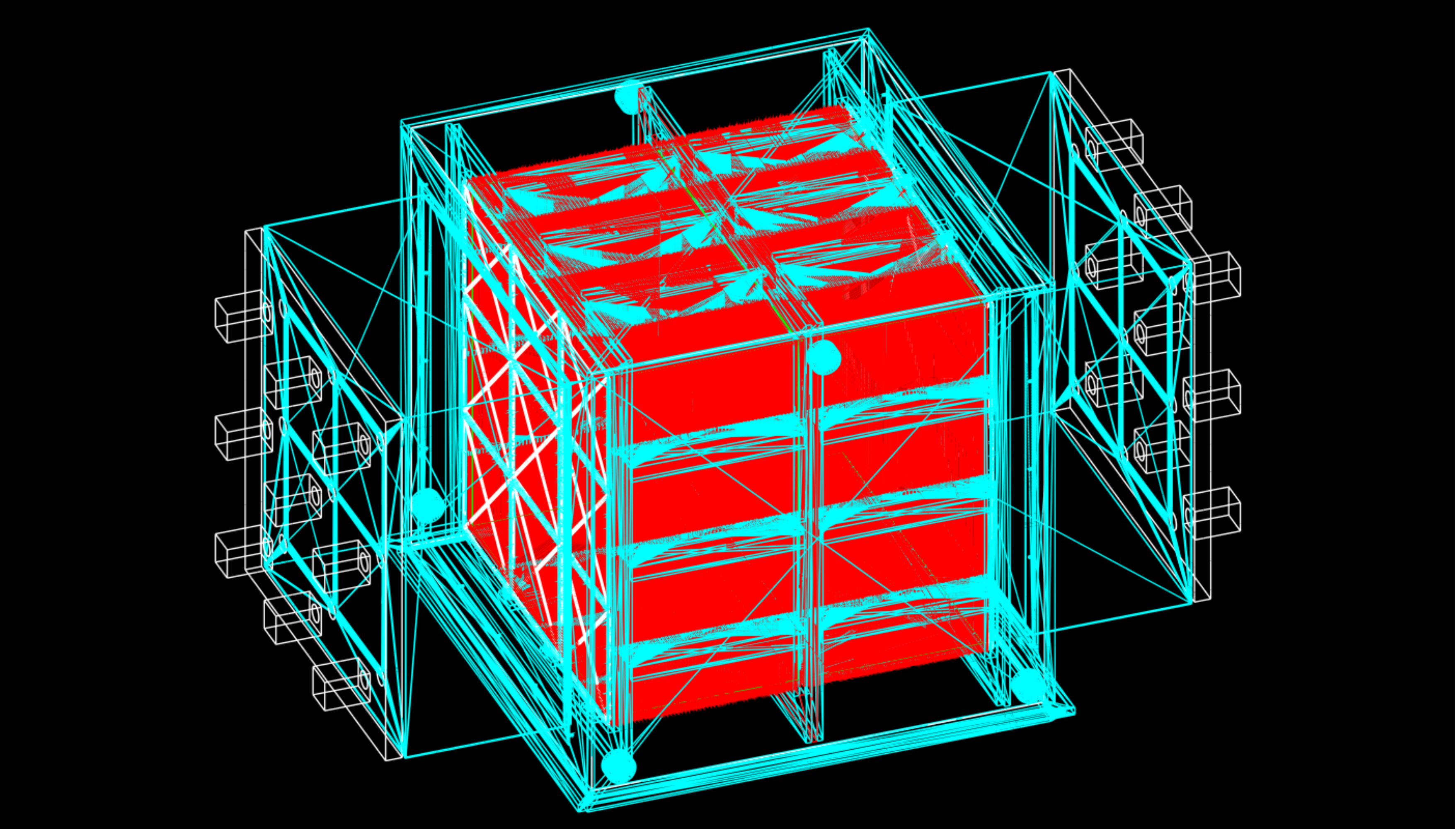}
\caption{\label{fig:geant4} The geometry of CYGNO detector implemented in the GEANT4 simulation.}
\end{figure}

A GEANT4~\cite{geant4} Monte Carlo (MC) simulation has been developed in order to study the expected background for CYGNO at LNGS and to optimize the geometry and the materials used for the shielding and for the general setup.
A detailed geometry of the detector has been implemented in the simulation (see Fig.\ref{fig:geant4}).

The MC simulation has been used to study several hypotheses for the external shielding, considering the effectiveness to screen from environmental photons and neutrons at LNGS, and taking also into account the radioactivity of shielding materials. 
The best option for CYGNO requirements has been identified in a concentric structure made of a inner layer of 5~cm of copper and water tanks outside, reaching a total of 2~m thickness on each side, top and floor.
This shielding provides an attenuation of about $10^{-7}$ for the photons flux and $5\cdot10^{-5}$ for the neutrons flux at LNGS.

\begin{figure}[htbp]
\centering 
\includegraphics[width=.40\textwidth]{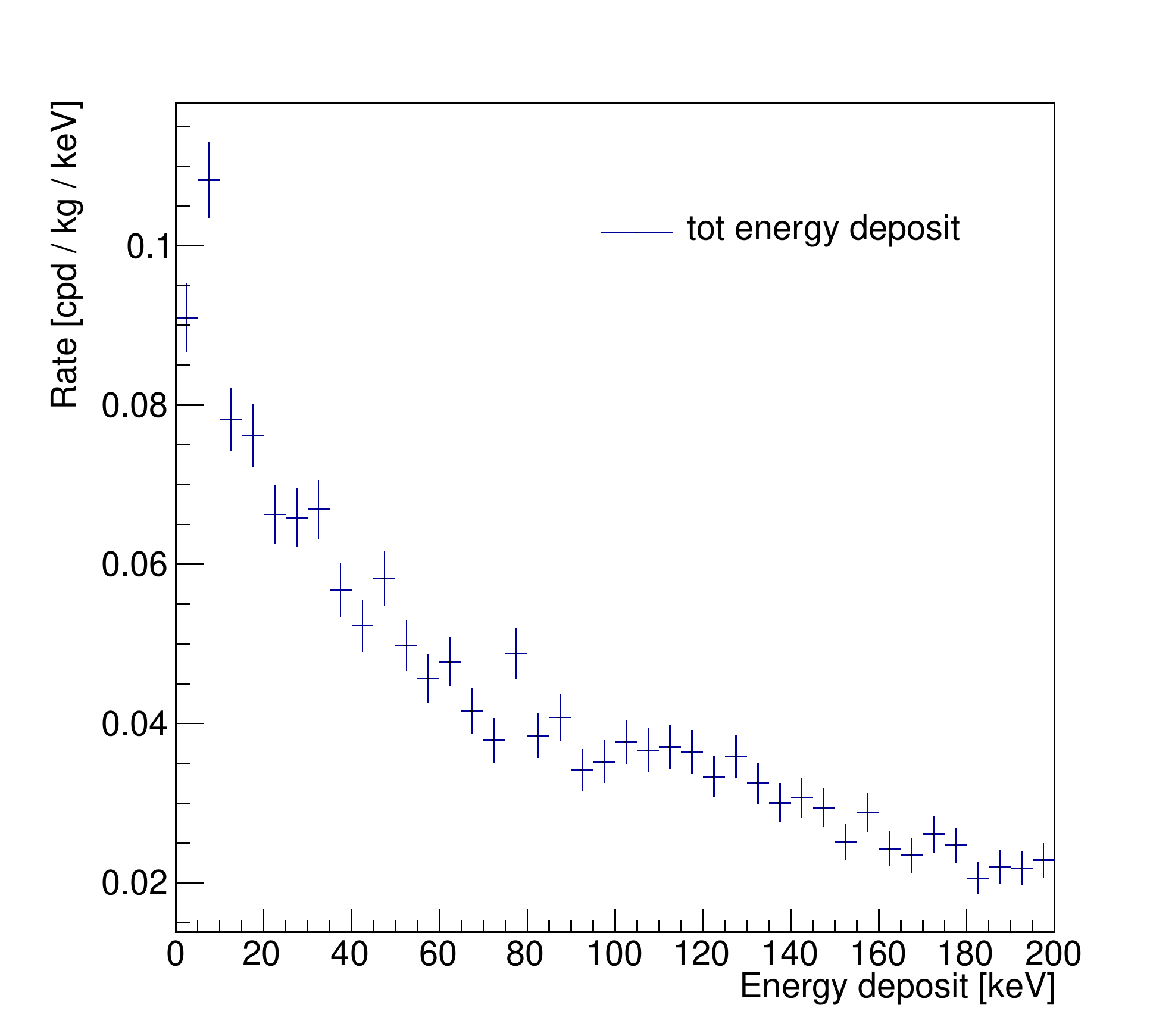}
\qquad
\includegraphics[width=.40\textwidth]{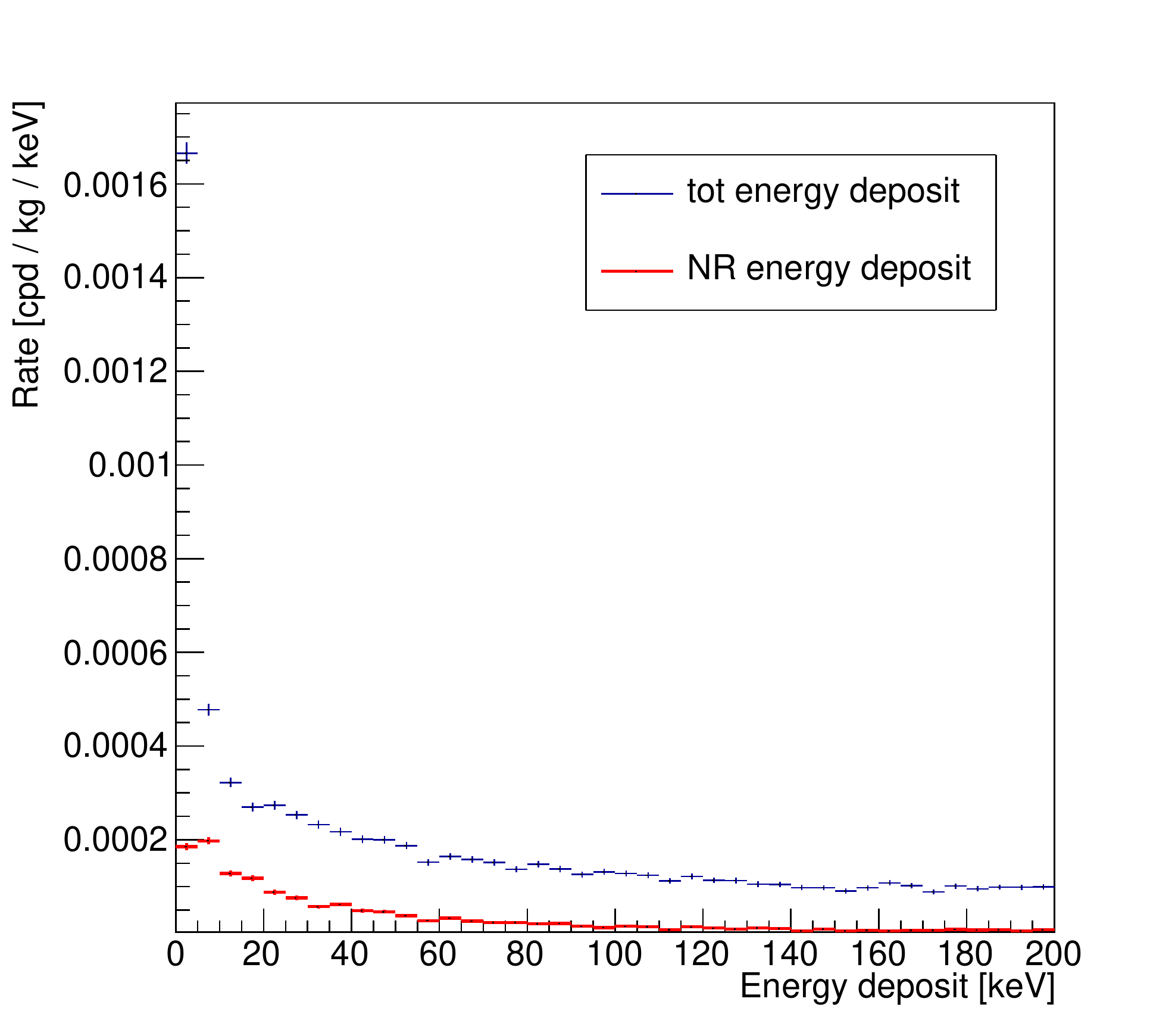}
  \caption{\label{fig:bkg} Expected background rate in counts/day/kg/keV in the CYGNO detector with a shielding made of 2~m of water and 5~cm of copper. Left plot shows the rate due to environmental photons at LNGS, right plot shows the total and the nuclear recoil rate due to neutrons at LNGS.}
\end{figure}

The expected background in the region of interest (ROI) at low energy ($<$20~keV) is of the order of $10^3$ counts/year for photons and $\sim1$ count/year for neutrons (see Fig.~\ref{fig:bkg}).

The goal of CYGNO is to reach a total background in the ROI of the order of $<10^{4}$~counts/year.
This value is motivated by the fact that, assuming a background rejection similar to other TPCs for dark matter search~\cite{DRIFT}, the expected background rate in the ROI is $<1$ count/year.
The estimate of the total background, of course, has to include not only the external background presented here, but also the internal background (i.e. radioactivity of the setup, especially the parts closer to the gas), that is expected to be the dominant one.
For this reason a careful screening of materials is in progress, and the MC simulation serves as an important tool to drive the choice of materials and optimize the geometry of the detector in order to minimize the radioactivity background.


\section{Expected sensitivity}
\label{sec:sensitivity}

\begin{figure}[htbp!]
\centering 
\includegraphics[width=.40\textwidth]{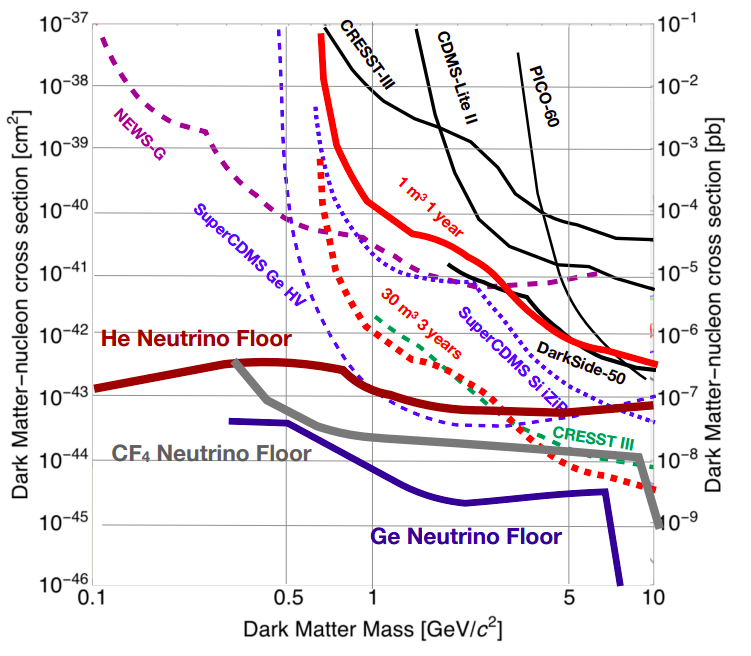}
\qquad
\includegraphics[width=.50\textwidth]{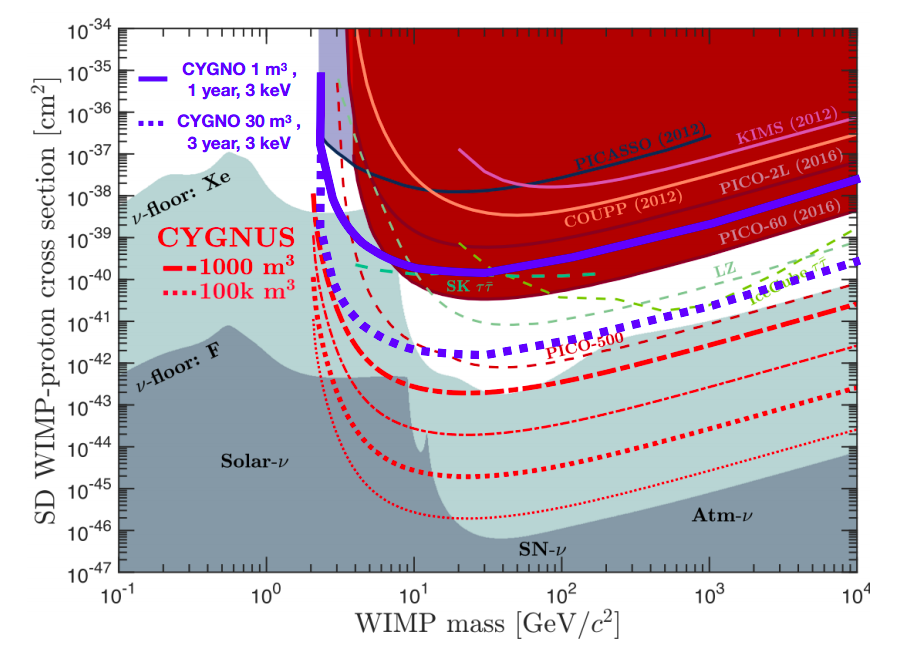}
  \caption{\label{fig:sensitivity} Expected spin-independent (left) and spin-dependent (right) 90\%C.L. exclusion limits for a 1~m$^3$ (30 m$^3$ dashed line) He CF$_4$ (80:20). A projection for 1000~m$^3$ is also shown for the spin-dependent case.}
\end{figure}

The projected sensitivity of CYGNO phase-1, with 1~m$^3$ volume and 1 year exposure, assuming zero background, is already competitive with the best limits of other experiments both in the spin-independent (SI) and in the spin-dependent (SD) scenario for low mass dark matter (see Fig.~\ref{fig:sensitivity}).
The phase-2, with 30~m$^3$ volume and three years exposure will also allow to lower the sensitivity and reach the neutrino floor in both scenarios.

CYGNO is part of a network of similar projects in several underground laboratories around the world (UK, US, Australia, Japan, China).
All these experiments will form a distributed observatory of solar and galactic particles detectors. 
The combination of results of all the detectors will improve the total sensitivity, reduce systematic uncertainties related to specificity of each experimental site, and increase the discovery potential.

\section{Conclusions}
\label{sec:conclusions}

Gaseous TPCs look very promising for the next generation of dark matter and neutrino detectors.
The development of the technology of optical sensors with high granularity and low noise makes the optical readout an interesting option for dark matter applications. 
The CYGNO project consists of a gaseous TPC using He:CF$_4$ gas mixture at atmospheric pressure, targeting 1~keV detection threshold, excellent background rejection and sensitivity to directionality for energies of few~keV.
The results of smaller prototypes show promising performances in terms of tracking resolution and energy threshold. 
CYGNO phase-1 with 1~m$^3$ of sensitive volume might already set a competitive limit on low mass dark matter and demonstrate the feasibility of a gaseous dark matter detector with directionality.
A larger gas volume, foreseen for the phase-2, is needed to reach the neutrino floor, and the sensitivity to directionality will become crucial to separate neutrinos from possible dark matter signal.
Directional detectors like CYGNO are needed to explore the phase space below neutrino cross sections, and, if dark matter is observed, to give an insight to the dark matter distribution in our galaxy.

%
%
%
%
%
%
%
\acknowledgments

This work was partially supported by the European Research Council (ERC) under the European Union's Horizon 2020 research and innovation program (grant agreement No 818744).

%


\end{document}